\documentclass[conference]{IEEEtran}
\IEEEoverridecommandlockouts

\newtheorem{Proposition}{\it Proposition}[section]

\makeatletter
\newcommand{\Rmnum}[1]{\expandafter\@slowromancap\romannumeral #1@}
\makeatother

\usepackage{amsmath, amssymb, amsfonts}
\usepackage{graphicx}
\usepackage{bm}
\usepackage{cuted}
\usepackage{multicol}
\usepackage{setspace}
\usepackage{subcaption}

\usepackage{epstopdf}
\usepackage{fancyhdr} % 添加页眉页脚
\usepackage{indentfirst}
\usepackage{enumerate}
\usepackage{stfloats}
\usepackage{bm}
\usepackage{multirow}
\usepackage{threeparttable}
\usepackage{CJK}
\usepackage[justification=centering]{caption}
\usepackage{makecell}
\usepackage{caption}
\usepackage{cases}
\usepackage{textcomp}
\usepackage{algpseudocode}
\usepackage{amsmath}
\usepackage{graphics}
\usepackage{epsfig}
\usepackage{cite}
\usepackage{color}
\usepackage{caption}

\usepackage{geometry}
\begin{document}
\captionsetup[figure]{labelformat={default},labelsep=period,name={Fig.}}
\title{\huge Integrated Sensing and Semantic Communication with Adaptive Source-Channel Coding\vspace{-0.5em}}

\author{
	\IEEEauthorblockN{Haotian Wang,  Dan Wang,  Xiaodong Xu,  Chuan Huang, Hao Chen and Nan Ma}
\vspace{-2em}

\thanks{
		%A conference version of this work \cite{7839497} was presented at IEEE GLOBECOM 2018.
		H. Wang, X. Xu, and N. Ma are with the State Key Laboratory of Networking and Switching Technology, Beijing University of Posts and Telecommunications, Beijing, China, 100876, and also with the Department of Broadband Communication, Peng Cheng Laboratory, Shenzhen, China, 518055. Emails: wanght@bupt.edu.cn, xuxiaodong@bupt.edu.cn, manan@bupt.edu.cn.
		
		D. Wang and H. Chen are with the Department of of Broadband Communications, Peng Cheng Laboratory, Shenzhen, China, 518055, and D. Wang is also with the Guangdong Provincial Key Laboratory of Future Networks of Intelligence, The Chinese University of Hong Kong, Shenzhen, 518172. Emails: wangd01@pcl.ac.cn, chenh03@pcl.ac.cn.
		
		Chuan Huang is with the Shenzhen Institute for Advanced Study, the University of Electronic Science and Technology of China, Shenzhen, 518110, China, and the Shenzhen Future Network of Intelligence Institute, Shenzhen, 518172, China (e-mail: huangch@uestc.edu.cn).

%		
%		\textit{Corresponding authors: Xiaodong Xu and Dan Wang.}
	}
}

%\thanks{fjvfjvjb
%}
\maketitle

\begin{abstract}
	Semantic communication  has emerged as a new paradigm to facilitate the performance of integrated sensing and communication  systems in 6G. However, most of the existing works mainly focus on sensing data compression to reduce the subsequent communication overheads, without considering the integrated transmission framework for both the SemCom and sensing tasks. 
	This paper proposes an adaptive source-channel coding and beamforming design framework for integrated sensing and SemCom  systems by jointly optimizing the coding rate for SemCom task and the transmit beamforming for both the SemCom and sensing tasks. Specifically, an end-to-end semantic distortion function is approximated by deriving an upper bound composing of source and channel coding induced components, and then a hybrid Cram\'{e}r-Rao bound (HCRB)  is also derived for target position under imperfect time synchronization. To facilitate the joint optimization, a distortion minimization problem is formulated by considering the HCRB threshold, channel uses, and power budget. Subsequently, an alternative optimization algorithm composed of successive convex approximation and fractional programming is proposed to address this problem by decoupling it into two subproblems for coding rate and beamforming designs, respectively. 
	Simulation results demonstrate that our proposed scheme outperforms the conventional deep joint source-channel coding -water filling-zero forcing  benchmark.
\end{abstract}
%\begin{IEEEkeywords}
%Semantic communication, adaptive source-channel coding (ASCC), integrated sensing and communication (ISAC),  Cram\'{e}r-Rao bound (CRB), beamforming design.
%\end{IEEEkeywords}

%\begin{IEEEkeywords}
%Interference channel (IC), amplify-and-forward (AF), full duplex (FD), rate region, concave-convex procedure (CCCP).
%\end{IEEEkeywords}
\newcounter{TempEqCnt}
\vspace{-0.5em}
\section{Introduction}
The sixth-generation (6G) communication systems are expected to support the extremely high-speed transmission and high-accuracy sensing demands of intelligent applications, such as autonomous driving, virtual/augmented reality (VR/AR), and immersive internet-of-things, which impose significant difficulties on the resource management for integrated sensing and communication (ISAC) in 6G \cite{11240213}. Semantic communication (SemCom), a new paradigm that leverages artificial intelligence (AI) to extract and transmit features of raw sensing data, has been proposed to reduce transmission overhead in ISAC systems \cite{10750351}. However, the inherent functional and operational differences between sensing and communications pose new design challenges for the integrated transmission framework to support both SemCom and sensing tasks.

%The development of emerging services such as autonomous driving and virtual/augmented reality (VR/AR) requires future 6G communication systems to not only possess high data rates, low latency, and massive connectivity, but also to achieve task-driven, context-aware information interaction in dynamic and diverse environments \cite{IEEE9170653}. However, traditional communication models aimed at bit-accurate transmission cannot effectively identify and extract task-critical information, leading to low efficiency in resource-limited, low SNR environments. In this context, semantic communication, as an important form of AI-enabled communication, transcends the traditional Shannon framework by shifting its focus from precise bit transmission to directly extracting, encoding, and transmitting the most task-relevant semantic information. This paradigm significantly improves resource utilization efficiency and is considered a key enabling technology for 6G.

In general, the SemCom-related research is divided into two categories, i.e., analog and digital cases \cite{8723589, yuan2025adaptivesourcechannelcodingmultiuser, 9398576, li2025adaptivesourcechannelcodingsemantic,10845799}. The authors in \cite{9398576} proposed a transformer-based deep joint source-channel coding (DJSCC) method in the analog SemCom system, which demonstrates superior performance in the low signal-to-noise-ratio (SNR) region compared to the
conventional communication. However, the semantic features are often encoded into analog symbols for transmission, which is incompatible with the digital communication systems in practice. To overcome this issue,
the authors in \cite{li2025adaptivesourcechannelcodingsemantic} investigated a digital DJSCC coding scheme to adaptively design the digital source-channel coding and superposition coding by minimizing the end-to-end (E2E) semantic distortion, which shows better performance on both reconstruction and classification tasks in low-SNR region. 

On the other hand, ISAC as one of the key technologies in 6G achieves both the sensing and communication functions by sharing the wireless resources and hardware platforms  \cite{9737357,10680280}. The combination of SemCom with ISAC systems has also attracted extensive attention from both the academic and industrial spheres \cite{10750351,10417099}. The authors in \cite{10750351} proposed a semantic-based sensing data feedback scheme for an autonomous aerial vehicle-assisted ISAC system, where a semantic database is utilized to extract high-level sensing features, thereby reducing feedback overhead and latency. Similarly, the authors in \cite{10417099} employed a deep neural network (DNN)-based encoder to compress multi-modal sensing data, and transmitted the extracted semantic features to the receiver, thereby significantly improving spectral efficiency.
However, most of the above mentioned existing studies have concentrated on  sensing data compression to reduce the transmission overhead, without considering the design of an integrated transmission framework that jointly serves SemCom and sensing tasks.

Motivated by the benefits of SemCom and ISAC in 6G, this paper proposes an adaptive source-channel coding (ASCC) framework for integrated sensing and SemCom (ISSC) systems. As the SemCom typically performs more effectively in low-SNR region, a larger proportion of resources could be allocated for sensing, thereby having the potential of bringing integrated performance gain.
%First, an E2E distortion is approximated by using a logistic regression model, and the hybrid Cram\'{e}r-Rao (HCRB) bound for target localization under imperfect time synchronization (TS) is also derived. Then, a distortion minimization problem is formulated by considering the CRB threshold, channel uses,  and power budget. Subsequently, an alternative optimization (AO) algorithm composing of successive convex approximation (SCA) and fractional programming (FP) is proposed to solve this problem. 
First, we approximate the end-to-end (E2E) distortion using a logistic regression model and derive the hybrid Cramér-Rao bound (HCRB) for target localization under imperfect time synchronization (TS). Then, a distortion minimization problem is formulated, constrained by the HCRB threshold, channel uses, and power. To solve this problem, we propose an alternating optimization (AO) algorithm combining successive convex approximation (SCA) and fractional programming (FP). Finally, simulation results demonstrate our analysis. 
\begin{figure*}[!t] 
	\setlength{\belowcaptionskip}{-20pt}
	\centering
	\includegraphics[width=0.6\textwidth]{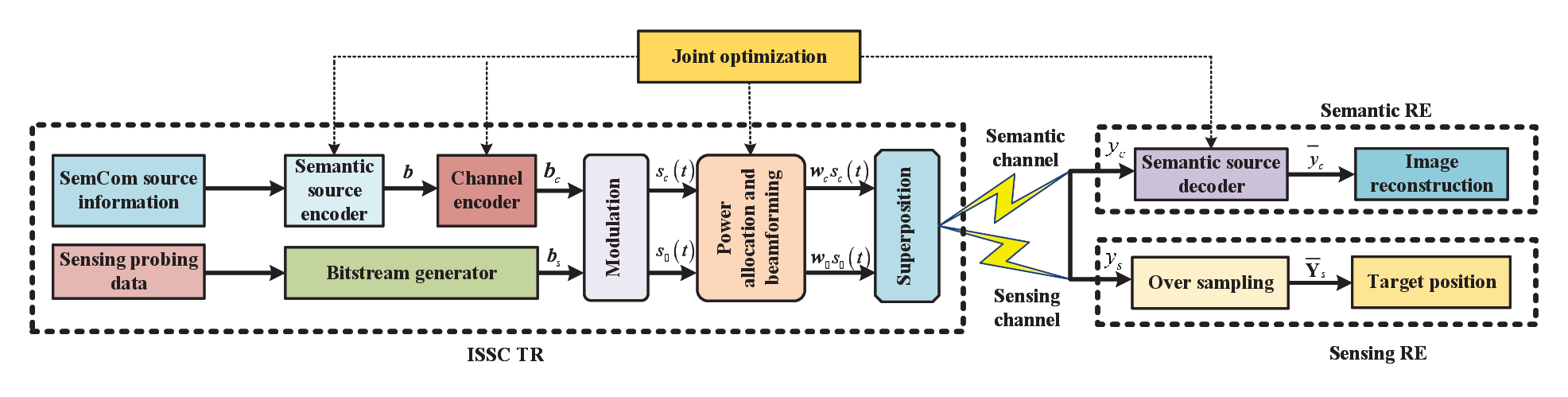} 
	\vspace{-2mm}   
	\caption{Framework of the ISSC System.}
	\label{fig:example}
\end{figure*}
\section{System Model}
In this paper, we consider a ISSC system consisting of one transmitter (TR), one target, and two receivers (REs), as shown in Fig.\ref{fig:example}. Here,  the TR and the sensing RE (SRE) are considered to be equipped with $N_T$ and $N_R$ antennas, respectively, while the SemCom RE (CRE) is considered to be equipped with a single antenna. The positions of  TR, target, CRE and  SRE are denoted as $\mathbf{p}_{b}=\left[b_x,b_y\right]^{T}$, $\mathbf{p}_{o}=\left[o_x,o_y\right]^{T}$, 	$\mathbf{p}_c=\left[c_x,c_y\right]^{T}$ and $\mathbf{p}_{r}=\left[r_x,r_y\right]^{T}$, respectively.
\subsection{Transmitter Design}
The ISSC TR integrates both SemCom and sensing functions.
For the semantic communication (SemCom) task, an $M$-dimensional observation vector~$\mathbf{j}$ is processed by a DNN-based semantic encoder~$f_{\varphi}$ to extract a feature vector~$\mathbf{z}$. 
This feature is then quantized as $\overline{\boldsymbol{z}}=\lfloor \boldsymbol{z}\rceil$, where $\lfloor\cdot \rceil$ denotes the uniform scalar quantization operation, and entropy-coded into a bitstream~$\mathbf{b}$ of total length~$B$ based on its probability mass function $\boldsymbol{P}_{\overline{\boldsymbol{z}}}\left(\overline{\boldsymbol{z}}\right)$. The expected source coding rate $R_s$ is calculated as $R_s\triangleq\mathbb{E}\left\{-\log_{2}\boldsymbol{P}_{\overline{\boldsymbol{z}}}\left(\overline{\boldsymbol{z}}\right)\right\}$.
For channel protection, $\mathbf{b}$ is partitioned into $I = \lfloor B/N_c \rfloor$ blocks and is transformed into $\mathbf{b}_c$ after channel encoding. 
Each $N_c$-bit block is modulated (e.g., QPSK, QAM, etc.) into an $L$-length symbol vector, and their concatenation forms the final signal~$\mathbf{s}_c \in \mathbb{C}^N$ with $N=IL$. 
Besides, the channel coding rate is calculated as $R_{c}=\frac{N_c}{L}$ and the average number of channel uses per source sample is defined as $E_a=\frac{R_s}{R_c}$. The metric measuring the average channel uses per dimension of $\mathbf{j}$ is defined as average bandwidth ratio (ABR), with $\text{ABR}=\frac{E_a}{M}$. 

For the sensing task, a bitstream generator produces the sensing bitstream $\mathbf{b}_s$ from the sensing probing data, which is then modulated---similar to the SemCom task---into a symbol vector $\mathbf{s}_0\in\mathbb{C}^{N}$.  
The SemCom symbols $\mathbf{s}_c$ and sensing symbols $\mathbf{s}_0$ are superimposed via beamforming to form the ISSC signal to perform tasks at REs.  
The TR transmits the ISSC signal to the target, and the SRE receives the reflected echo to estimate the target’s position.  
Specifically, the ISSC signal transmitted by the TR at the $t$‑th time slot is given by:
\begin{equation}\label{1}
	\mathbf{x}{\left(t\right)}=\sum_{u\in\mathcal{U}}\mathbf{w}_{u}(t){s}_{u}{\left(t\right)}, t\in\mathcal{T},
\end{equation}
where $\mathcal{U}=\{c,0\}$ is the set of subscripts, with $c$ and $0$ being for SemCom and sensing tasks, respectively; $\mathbf{w}_{u}(t)\in\mathbb{C}^{N_{T}\times 1}$ denotes the beamforming vector for $u$-th task; $\mathcal{T}=\{1,\cdots,T\}$, with $T$ being the number of total time slots; The continuous-time signal is formed by  ${s}_{u}\left(t\right)=\sum_{n\in{N}}{s}_{u}\left[n\right]g\left(t-\left(n-1\right)\Delta t\right)$. Here, $s_u[n]$ are symbols from a constellation~$\mathcal{S}$ with zero mean and unit power (i.e., $\mathbb{E}\{s_u\} = 0$, $\mathbb{E}\{|s_u|^2\} = 1$), and $g(t)$ is a unit-energy pulse shaping filter satisfying $\frac{1}{\Delta t} \int_0^{\Delta t} |g(t)|^2 dt = 1$. 
The total transmission power at the TR is constrained by a budget~$P_T$:
\begin{equation}
	\mathbb{E}\{\|\mathbf{x}(t)\|^2\} = \mathrm{Tr}(\mathbf{W}_c + \mathbf{W}_0) = \mathrm{Tr}(\boldsymbol{\Sigma}) \leq P_T,
\end{equation}
where $\mathbf{W}_{c}=\mathbf{w}_{c}\mathbf{w}_{c}^{H}\in\mathbb{C}^{N_{T}\times N_{T}}$ and $\mathbf{W}_{0}=\mathbf{w}_{0}\mathbf{w}_{0}^{H}\in\mathbb{C}^{N_{T}\times N_{T}}$ represent the covariance matrices for SemCom and sensing tasks, respectively; and $P_T$ is the power budget at TR.
\vspace{-2em}
\subsection{Receiver Designs} 
%In this paper, we consider an ISSC system consisting of one  TR, one target, and two REs. The TR  transmits ISSC signal to the target and  REs. The first RE receives the reflected signal from target for sensing task and the second one directly  receives  signal from TR for semantic communication (semcom) task, e.g., image reconstruction. Here,  the TR and the sensing RE (SRE) are considered to b equipped with $N_T$ antennas and $N_R$ antennas, respectively, and the semcom RE (CRE) is considered to be equipped with single antenna. The positions of  TR, target, and  SRE are denoted as $\mathbf{p}_{b}=\left[b_x,b_y\right]^{T}$, $\mathbf{p}_{o}=\left[o_x,o_y\right]^{T}$, 	and $\mathbf{p}_{r}=\left[r_x,r_y\right]^{T}$, respectively.
\subsubsection{Designs for CRE}
For the SemCom task, the TR directly transmits signal ${\bf x}(t)$ to CRE, and then the received signal $ y_{c}\left(t\right)$ at the CRE is given as
\begin{align}\label{equ: received signal1}
	{y}_{c}\left(t\right)&=\mathbf{h}_{c}^{H}\mathbf{x}\left(t\right)+n\left(t\right)\nonumber\\
	&=\mathbf{h}_{c}^{H}\mathbf{w}_{c}{s}_{c}\left(t\right)+\mathbf{h}_{c}^{H} \mathbf{w}_{0}{s}_{0}\left(
	t \right)+n\left(t\right),
\end{align}
where $\mathbf{h}_{c}=\sqrt{\eta}_c\boldsymbol{\alpha}\left(\varphi\right)\in\mathbb{C}^{N_{T}\times 1}$ represents the channel coefficient of the direct link from TR to CRE and $\boldsymbol{\alpha}\left(\cdot\right)=\left[
1, e^{j2\pi\frac{d_a}{\lambda}sin\left(\cdot\right)},\cdots,e^{j2\pi\left(N_{T}-1\right)\frac{d_a}{\lambda}sin\left(\cdot\right)}\right]^{T}$ is the array steering vector with $\lambda$ being the carrier wavelength and $d_a$ being the distance between any two adjacent antennas. Besides,  $\eta_c=d_c^{-\epsilon}$ represents the path loss; $d_c$ denotes the distance between TR and CRE; $\varphi=\mathbf{arctan}\left(\frac{c_y-b_y}{c_x-b_x}\right)+\mathbb{I}_{\{c_x<b_x\}}\pi$ denotes the direction of departure (DoD) from TR with $\mathbb{I}_{\{\cdot\}}$ being the indicator function, $\epsilon$ denotes the path loss exponent; and  $n\left(t\right)$ is the  received circularly symmetric complex Gaussian (CSCG) noise satisfying $n\left(t\right)\sim\mathcal{CN}\left(0,\sigma_{n}^{2}\right)$.
Then, the decoded signal $\overline{y}_{c}$ is input into the DNN-based function $\boldsymbol{g}_{\nu}\left(\overline{y}_{c}\right)$  to recover the transmitted data $\mathbf{j}$, with $\overline{\mathbf{j}}=\boldsymbol{g}_{\nu}\left(\overline{y}_{c}\right)$ representing the recovered data. Notably, the average bit error probability during transmission with random coding is approximated as \cite{5452208}
\begin{equation}\label{7}
	\rho_b=\frac{Q\left(\ln 2\frac{\sqrt{L}\left(C\left(\gamma\right)-R_c\right)}{\sqrt{1-\frac{1}{\left(1+\gamma\right)^2}}}\right)}{R_c L},
\end{equation}
where  $Q\left(x\right)=\frac{1}{2\pi}\int_{x}^{\infty}e^{-\frac{t^2}{2}}dt$, and $C\left(\gamma\right)=\text{log}_{2}\left(1+\gamma\right)$ denotes the channel capacity, with $\gamma=\frac{\left|\mathbf{h}_c^{H}\mathbf{w}_c\right|^2}{\left|\mathbf{h}_c^{H}\mathbf{w}_0\right|^2+\sigma_{n}^2}$ representing   the signal-to-interference-plus-noise ratio (SINR) at the CRE.
\subsubsection{Designs for SRE}
For the sensing task, the TR transmits signal ${\bf x}(t)$ to the target and the signal is then reflected to SRE  to estimate the target's position $\mathbf{p}_o=\left[o_x,o_y\right]^{T}$. Thus, the received signal at the  SRE is given as 
\begin{align}\label{9}
	\mathbf{y}_{s}\left(t\right)=\mathbf{H}_{0}\mathbf{x}\left(t-\tau\right)+\mathbf{n}_s\left(t\right),	
\end{align} 
where $\mathbf{H}_{0}=\beta_{0}\boldsymbol{\alpha}\left(\phi\right)\boldsymbol{\alpha}\left(\theta\right)^{H}\in\mathbb{C}^{N_{R}\times N_{T}}$, with $\beta_{0}$ representing the reflection cofficient incorporating the effects of the radar cross section (RCS) and path loss, and $\phi$ and $\theta$ being the DoD from the TR to the target and the direction of arrival (DoA) from the target to the SRE, respectively; $\tau$ is the transmission delay of TR-target-SRE link; $\mathbf{n}_s\left(t\right)$ is the  received CSCG noise with $\mathbf{n}_s\left(t\right)\sim\mathcal{CN}\left(0,\sigma_{s}^{2}\mathbf{I}_{N_R}\right)$.  The DoD $\phi$ and DoA $\theta$  depend on the positions of TR, target, and SRE, i.e., $\theta=\mathbf{arctan}\left(\frac{o_y-b_y}{o_x-b_x}\right)+\mathbb{I}_{\{o_x<b_x\}}\pi$ and $\phi=\mathbf{arctan}\left(\frac{o_y-r_y}{o_x-r_x}\right)+\mathbb{I}_{\{o_x<r_x\}}\pi$.
Here, we consider imperfect time synchronization (TS), which significantly impacts positioning accuracy\cite{10684491}. The TR-SRE transmission delay is modeled as $\tau=\tau_{t,r}+\triangle\tau_{t,r}$, 
where $\tau_{t,r}$ is the transmission delay, and $\Delta	\tau_{t,r}$ is the random TS error, satisfying $\triangle{\tau_{t,r}}\sim\mathcal{CN}\left(0,\sigma_{\triangle}^{2}\right)$. 
The error variance~$\sigma_\Delta^2$ is assumed to be known at the SRE. 
Additionally, the SRE is assumed to be capable of mitigating stationary clutter and interference from the TR-SRE link~\cite{10684491}.
\section{Performance analysis }
In this section, we model SemCom and sensing performance by the E2E distortion and  HCRB under imperfect TS, respectively.
\subsection{E2E Distortion Model for SemCom Task}
The distortion function $D_{o}$ is modeled as the sum of source and channel distortions, i.e.,  
\begin{equation}\label{11}
	D_{o}\approx D_{o}^{s}\left(R_s\right)+D_{o}^{c}\left(R_s, \rho_{b} \right), 
\end{equation}
where $D_o^s(R_s)$ and $D_o^c(R_s, \rho_b)$ are the source and channel distortions, respectively. 
Obtaining a closed-form expression for the channel distortion~$D_o^c(R_s, \rho_b)$ is intractable due to its complex dependence on the DNN's Lipschitz constant and channel coding scheme. 
Therefore, we adopt a data regression approach: we first pre-train multiple data recovery models and then fit their performance to derive a closed-form expression for the total distortion~$D_o$. 
The details are given as follows:
\subsubsection{Designs of Source Feature Extraction Module}
$G$ hyper-prior based DNN models \cite{Balle2018Variational}  are employed to acquire the feature function $\boldsymbol{f}_{\varphi}$ and the recovery function $\boldsymbol{g}_{\nu}$, where the set of source coding rate corresponding to each model is denoted as $\mathcal{R}_{s}=\left\{R_{s,1},...,R_{s,G}\right\}$. Moreover, each model is trained under the ideal error-free transmission conditions.
\subsubsection{Distortion measures}
Without loss of generality, the widely studied image recovery is considered as an example. The well-known mean square error (MSE)  is utilized to measure the average distortion, i.e, $	D_{o}=E\left(d_o\left(\mathbf{j},\hat{\mathbf{j}}\right)\right)$, with $d_o\left(\mathbf{j},\hat{\mathbf{j}}\right)=\frac{1}{M}\left|\left|\mathbf{j}-\hat{\mathbf{
		j}}\right|\right|_2^2$. 
\subsubsection{E2E distortion approximation}
\begin{figure}[h]
 	\setlength{\belowcaptionskip}{-20pt}
	\setlength{\abovecaptionskip}{0.2cm}
	\centering
	\vspace{-1em}
	\includegraphics[width=1.5in]{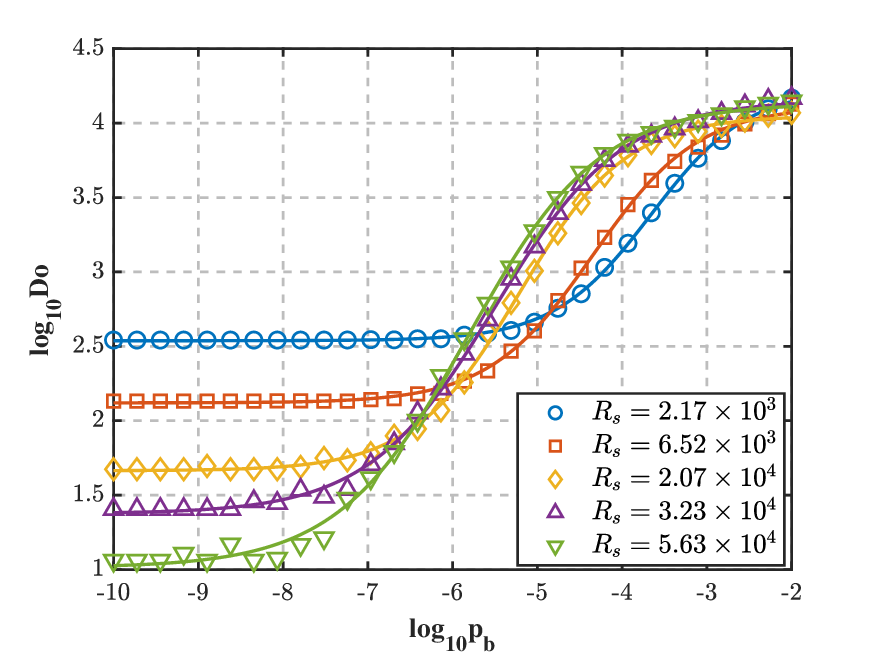}
	\caption{E2E distortion as a function of $\rho_b$ under different $R_s$.}
	\label{2}

\end{figure}

To approximate the E2E distortion $D_o$, we first simulate  distortion by randomly flipping bits according to a specific BER $\rho_b$ on the Caltech-UCSD Birds 200 (CUB-200-1011) data-set.  Then we calculate the E2E distortion $\mathbf{D}_o=\left[
D_{o,1},...,D_{o,G}\right]$ corresponding to $G$ pre-trained data encoder-decoder pairs. As depicted in Fig.\ref{2}, the relationship between the resulting distortion $\log_{10} D_o$ and $\log_{10} \rho_{b}$ exhibits a distinct S-shaped curve, which allows for deriving a closed-form expression via the data regression (DR) method, i.e.,
\begin{equation}
\begin{aligned}\label{13}
		\log_{10} D_{o} 
		\triangleq \hat{D_{o}^{s}}\left(R_s\right)+\frac{\hat{D_{o}^{c}}\left(R_s\right)}{1+e^{-E_{1}^{o}\left(R_s\right)\left(\log_{10} \rho_{b}-E_{2}^{o}\left(R_s\right)\right)}},
\end{aligned}
\end{equation}
where $\hat{D_{o}^{s}}\left(R_s\right),\hat{D_{o}^{c}}\left(R_s\right), -E_{1}^{o}\left(R_s\right), E_{2}^{o}\left(R_s\right)$ are all fitting parameters that need to be determined.
Fig.~\ref{2} illustrates a key trade-off: while a higher source rate~$R_s$ reduces source distortion, it increases the sensitivity of the total semantic distortion~$D_o$ to the bit error rate (BER)~$\rho_b$. 
Minimizing~$D_o$ thus requires optimizing the balance between~$R_s$ and~$\rho_b$.
\begin{figure*}[b] 
	\hrulefill  
	\vspace{-0.5em}  
	\begin{equation}  
		\label{eq:hcrlb}
		\text{HCRB}(\boldsymbol{\mu}) =
		\underbrace{\Bigl(\mathbf{J}_{D}^{\mathbf{c}_{1}}\Bigr)^{-1}}_{\text{CRB}\left(\mathbf{p}_o\right)}
		+\underbrace{\Bigl(\mathbf{J}_{D}^{\mathbf{c}_{1} }\Bigr)^{-1} 
			\left(\mathbf{J}_{D}^{\mathbf{c}_{2} }\right)
			\biggl[ \mathbf{J}^{c_3} 
			- \Bigl(\mathbf{J}_{D}^{\mathbf{c}_{2} }\Bigr)^{{T}}
			\Bigl(\mathbf{J}_{D}^{\mathbf{c}_{1} }\Bigr)^{-1}
			\mathbf{J}_{D}^{\mathbf{c}_{2}} \biggr]^{-1}  
			\Bigl(\mathbf{J}_{D}^{\mathbf{c}_{2} }\Bigr)^{{T}}
			\Bigl(\mathbf{J}_{D}^{\mathbf{c}_{1} }\Bigr)^{-1}}_{\mathrm{additional\ error\ caused \ by\ TS\ error}}.
			\tag{13}
	\end{equation}
	\vspace{-2em}
\end{figure*}
\subsection{HCRB Derivation for Sensing Task}
For the sensing task, the unknown estimation parameter vector is denoted as $\boldsymbol{\mu}=\left[\mathbf{p}_o,\triangle \tau_{t,r}\right]\in\mathbb{R}^{3\times1}$, which is estimated at the SRE. Then, the HCRB, a theoretical limit on the variance of any estimation parameter, is adopted as the performance metric in our considered ISSC system. 

To simplify the analysis, the continuous received signal $\mathbf{y}_{s}\left(t\right)$ is 
sampled with interval $\triangle t$, and then $N$ independent observations $\overline{\mathbf{Y}}_s=\left[\overline{\mathbf y}_s\left(t_1\right),\overline{\mathbf y}_s\left(t_2\right),...,\overline{\mathbf y}_s\left(t_N\right)\right]$ are obtained at the SRE, with $t_n=n\triangle t$ and $n\in\mathcal{N}=\left\{1,2,\cdots,N\right\}$. To investigate the impact of TS error on target positioning, the sampled signal is further mapped into frequency domain via DFT, and then the signal on frequency domain corresponding to the $k$-th $\left(k\in\left[0,1,\ldots, N-1\right]\right)$ sample is given as
\begin{equation}\label{15}
	\breve{\mathbf{y}}_{s}\left(w_k\right)=\mathbf{H}_{0}\breve{\mathbf{x}}\left(w_k\right)+\breve{\mathbf{n}}_{s}\left(
	w_k\right),
\end{equation} 
where $\check{\mathbf{y}}_s(w_k)$ is the $k$-th frequency-domain sample of the received signal, with $\check{\mathbf{x}}(w_k) = e^{-j w_k\tau} (\mathbf{w}_c \check{s}_c(w_k) + \mathbf{w}_0 \check{s}_0(w_k))$, $w_k = \frac{2\pi k}{N\Delta t}$, and $\breve{s}_c\left(w_k\right)=\sum_{n=1}^{N}s_{c}\left[n\right]e^{-j2\pi k\frac{n}{N}}$, $\breve{s}_0\left(w_k\right)=\sum_{n=1}^{N}s_{0}\left[n\right]e^{-j2\pi k\frac{n}{N}}$ and $\breve{\mathbf{n}}_{s}\left(
w_k\right)$ respectively denoting the DFT of $s_c\left(t_k\right)$, $s_0\left(t_k\right)$ and $\mathbf{n}_{s}\left(t_k\right)$, respectively.
The system model in matrix form is $\check{\mathbf{Y}}_s = \mathbf{H}_0 \check{\mathbf{X}} + \check{\mathbf{N}}_s\in\mathbb{C}^{N_{R}\times N}$.
To simplify the analysis, the frequency domain signal $\breve{\mathbf{Y}}_s$ is further rewritten as a column vector form, i.e., $\breve{\mathbf{y}}_{s}^{'}=	\boldsymbol{\chi}+\breve{\mathbf{n}}_{s}^{'}$, where $\breve{\mathbf{y}}_{s}^{'}=\text{vec}\left(\breve{\mathbf{Y}}_s\right)$, $\boldsymbol{\chi}=\text{vec}\left(\mathbf{H}_{0}\breve{\mathbf{X}}\right)$ and $\breve{\mathbf{n}}_{s}^{'}=\text{vec}\left(\breve{\mathbf{N}}_s\right)\sim\mathcal{CN}\left(0,\sigma_{s}^{2}\mathbf{I}_{N_{R}N}\right)$.
As the TS error $\triangle
\tau_{t,r}$ is a random variable, the HCRB, which  describes a lower bound of estimations of deterministic and random variables, is utilized as the sensing performance metric in our considered ISSC system. Compared with conventional CRB, the HCRB achieves a precise lower bound with additional prior information \cite{6877741}. The expression of the hybrid FIM is $\mathbf{J}\left(\boldsymbol{\mu}\right)=\mathbf{J}_{D}\left(\boldsymbol{\mu}\right)+\mathbf{J}_{B}\left(\boldsymbol{\mu}\right)$,
where $\mathbf{J}_{D}\left(\boldsymbol{\mu}\right)$ and $\mathbf{J}_{B}\left(\boldsymbol{\mu}\right)$ respectively denote the observed FIM and the prior FIM. The expression of  $\mathbf{J}_{D}\left(\boldsymbol{\mu}\right)$ is given as $\mathbf{J}_{D}\left(\boldsymbol{\mu}\right)=\mathbb{E}_{\boldsymbol{\mu}}\left\{\overline{\mathbf{J}}_D\left(\boldsymbol{\mu}\right)\right\}$, where
\begin{align}\label{18}
	\overline{\mathbf{J}}_D\left(\boldsymbol{\mu}\right)=-\mathbb{E}_{\breve{\mathbf{y}}_{s}^{'}|\boldsymbol{\mu}}\left(\frac{\partial^{2}\text{ln}\hat{\boldsymbol{f}}\left(\breve{\mathbf{y}}_{s}^{'}|\boldsymbol{\mu}\right)}{\partial\boldsymbol{\mu}\partial\boldsymbol{\mu}^T}\right),
\end{align}
with $\hat{\boldsymbol{f}}\left(\breve{\mathbf{y}}_{s}^{'}|\boldsymbol{\mu}\right)$ denoting the conditional probability density function of the observation $\breve{\mathbf{y}}_{s}^{'}$, i.e.,
\begin{equation}\label{19}
	\hat{\boldsymbol{f}}\left(\breve{\mathbf{y}}_{s}^{'}|\boldsymbol{\mu}\right)=\frac{\text{exp}\left\{-\frac{1}{2}\overline{y}_s^{T}\left(\sigma_{s}^{2}\mathbf{I}_{N_{R}N}\right)^{-1}\overline{y}_s\right\}}{\left(2\pi\left| \sigma_{s}^{2}\mathbf{I}_{N_{R}N}\right|\right)^{\frac{N_R}{2}}},
\end{equation}
and $\overline{y}_s=\left(\breve{\mathbf{y}}_{s}^{'}-\boldsymbol{\chi}\right)$. The HCRB is tight, which has been proved in \cite{10684491}. Therefore, we have $\mathbb{E}_{\boldsymbol{\mu}}\left\{\overline{\mathbf{J}}_D\left(\boldsymbol{\mu}\right)\right\}= \overline{\mathbf{J}}_D\left(\boldsymbol{\mu}\right)$. In addition, the prior FIM $\mathbf{J}_{B}\left(
\boldsymbol{\mu}\right)$ is $\mathbf{J}_{B}\left(\boldsymbol{\mu}\right)=-\mathbb{E}_{\boldsymbol{\mu}}\left(\frac{\partial^{2}\text{ln}\rho\left(\Delta \tau_{t,r}\right)}{\partial\boldsymbol{\mu}\partial\boldsymbol{\mu}^T}\right)$,
where $\rho\left(\Delta \tau_{t,r}\right)$ is the prior distribution of $\Delta \tau_{t,r}$, satisfying $\rho(\Delta	\tau_{t,r}) = \exp(-\Delta	\tau_{t,r}^2 / (2\sigma_\Delta^2)) / (\sqrt{2\pi}\sigma_\Delta)$.
Based on above analysis, we have the following proposition to derive FIM $\mathbf{J}\left(\boldsymbol{\mu}\right)$.
\begin{Proposition}
	FIM $\mathbf{J}\left(\boldsymbol{\mu}\right)$ is equivalently transformed into a block matrix form by taking the first-order  derivatives with respect to $\boldsymbol{\mu}$, i.e., 
	\begin{equation}\label{22}
		\mathbf{J}\left(\boldsymbol{\mu}\right)=
		\begin{bmatrix}
			\mathbf{J}_D^{\mathbf{c}_1}&\mathbf{J}_D^{\mathbf{c}_2} \\
			\left(\mathbf{J}_D^{\mathbf{c}_2}\right)^T&\mathbf{J}_D^{c_3}			
		\end{bmatrix}+\mathbf{J}_B\left(\boldsymbol{\mu}\right),
	\end{equation}
where $\mathbf{c}_1=\mathbf{p}_o\mathbf{p}_o$, $\mathbf{c}_2=\mathbf{p}_o\Delta \tau_{t,r}$, $c_3=\Delta \tau_{t,r}\Delta \tau_{t,r}$ and each block matrix in $\mathbf{J}_D\left(\boldsymbol{\mu}\right)$ is derived as
	\begin{align}\label{23}
		\mathbf{J}_D^{\mathbf{c}_1}=
		\begin{bmatrix}
			J_D^1&J_D^2 \\
			J_D^2&J_D^3
		\end{bmatrix},		                   \mathbf{J}_D^{\mathbf{c}_2}=\begin{bmatrix}
			J_D^4\\J_D^5
		\end{bmatrix},
		\mathbf{J}_D^{c_3}=J_D^6,
	\end{align}
	with the  expression of $J_D^j$  defining as $J_D^j=\frac{2}{\sigma_{s}^2}\mathcal{R}\left\{\text{Tr}\left(\boldsymbol{\zeta}_{j}\boldsymbol{\Sigma}\right)\right\}$,   $j\in\boldsymbol{\mathcal{J}}=\left\{1,...,6\right\}$,  $\boldsymbol{\Sigma}=\mathbf{W}_c+\mathbf{W}_0$, and each $\boldsymbol{\zeta}_j$ being $\boldsymbol{\zeta}_{1} =\sum_{n=1}^{N} \left( \dot{\mathbf{H}}_{0}^{n} \right)^{H}  \dot{\mathbf{H}}_{0}^{n}$, $\boldsymbol{\zeta}_{2} =\sum_{n=1}^{N}\left( \dot{\mathbf{H}}_{0}^{n} \right)^{H}  \ddot{\mathbf{H}}_{0}^{n}$, $\boldsymbol{\zeta}_{3} =\sum_{n=1}^{N} \left( \ddot{\mathbf{H}}_{0}^{n} \right)^{H}  \ddot{\mathbf{H}}_{0}^{n}$, $\boldsymbol{\zeta}_{4}=\sum_{n=1}^{N} -j w_{n}\left( \dot{\mathbf{H}}_{0}^{n} \right)^{H} \mathbf{H}_{0}$, $	\boldsymbol{\zeta}_{5}=\sum_{n=1}^{N} -j w_{n}\left( \ddot{\mathbf{H}}_{0}^{n} \right)^{H} \mathbf{H}_{0}$ and $\boldsymbol{\zeta}_{6} =\sum_{n=1}^{N} w_{n}^{2}\left( \mathbf{H}_{0} \right)^{H} \mathbf{H}_{0}$.
	Here, $\dot{\mathbf{H}}_{0}^{n}=\mathbf{H}_{0}^{'}-jw_{n}\tau^{'}\mathbf{H}_{0}$ , $\ddot{\mathbf{
			H}}_0^n=\mathbf{H}_{0}^{''}-jw_{n}\tau^{''}\mathbf{H}_0$ and $\mathbf{H}_{0}^{'}$ and $\mathbf{H}_{0}^{''}$ are the first-order derivative of $\mathbf{H}_0$ with respect to $o_x$ and $o_y$, respectively. Moreover, $\mathbf{J}_B\left(\boldsymbol{\mu}\right)=\begin{bmatrix}
		\mathbf{0}_{2\times2}&\mathbf{0}_{2\times1} \\
		\mathbf{0}_{1\times2}&\frac{1}{{\sigma_{\Delta}}^2}
	\end{bmatrix}$.
\end{Proposition}

Based on proposition 3.1, the HCRB matrix ${\bf J}^{-1}$, which is defined as the inverse of FIM, is derived as (\ref{eq:hcrlb}),
where $\mathbf{J}^{c_3}=\mathbf{J}_{D}^{c_3}+\frac{1}{\sigma_{\triangle}^2}$ .
Obviously, the presence of TS errors degrades estimation performance by introducing an additional penalty in  HCRB for the target position $\mathbf{p}_o$.

\section{ISSC Problem Formulation}
In this section, we first formulate the E2E distortion minimization problem. Then, an AO algorithm composed of SCA and FP  is proposed to address this problem.  
\subsection{Problem Formulation}
Following the analysis in section \uppercase\expandafter{\romannumeral3}, an E2E distortion minimization problem, subject to the power budget, channel uses, and HCRB threshold,  is formulated as 
\begin{align}
	\setcounter{equation}{13}
	\left(\text{P4.1}\right)\mathop{\text{min}}\limits_{R_s\in\mathcal{R}_s, R_c,\mathbf{w}_0,\mathbf{w}_c} ~~~~&~D_o \label{26}\\
	\text{s.t.}~~~~~~~~~~~&~\text{Tr}\left\{\text{HCRB}\left(\boldsymbol{\mu}\right)\right\} \leq\Pi \label{27}, \\
	~~~~&~\frac{R_s}{R_c}\leq E_a^{\text{max}},\\
	~~~~&~ 0 < R_c\leq C\left(\gamma\right) \label{28},\\
	~~~~&~\mathbf{Tr}\left(\boldsymbol{\Sigma}
	\right)\leq\ P_{T} \label{29},
\end{align}	
where $\Pi$ and $E_a^{\text{max}}$ denote the  HCRB threshold and maximum channel uses, respectively. Problem~(P4.1) is a mixed-variable optimization, since the source rate $R_s$, which is determined by a pre-trained DNN, is a discrete variable from the set $\mathcal{R}_s = \{R_{s,1}, \dots, R_{s,G}\}$. To solve this problem, we first fix $R_s$ and  transform Problem~(P4.1) into 
\begin{align}
	\left(\text{P4.2}\right)\mathop{\text{min}}\limits_{R_c,\mathbf{w}_0, \mathbf{w}_c} ~~~~&~10^{\left(\hat{D_{o}^{s}}\left(R_s\right)+\frac{\hat{D_{o}^{c}}\left(R_s\right)}{1+e^{-E_{1}^{o}\left(R_s\right)\left(\log_{10} \rho_{b}-E_{2}^{o}\left(R_s\right)\right)}}\right)} \label{30}\\
	\text{s.t.} 
	~~~~~~&~\frac{R_s}{E_a^{\text{max}}}\leq R_c\leq C\left(\gamma\right), \\
	~~~~&~(15), (18)\label{31}, 
\end{align}	 
where (19) is the closed-form expression of (14) and (20) is obtained from (16) and (17). Then, the optimal $R_s$ corresponding to the optimal DNN model is acquired by comparing the minimum value of (14) corresponding to different $R_s \in \mathcal{R}_s$. However, it is obvious that Problem~(P4.2) is still non-convex due to its fractional objective function in (19)  and the intractable HCRB expression in (15). 
Hence, we decompose it  into two sub-problems for channel coding rate  and  beamforming optimization. 
\subsubsection{Coding Rate Optimization}
For simplicity, two logarithmic auxiliary variables $\hat{\rho}_b=\text{log}_{10}\rho_{b}$ and $e^{\nu}=\frac{\hat{D_{o}^{c}}\left(R_s\right)}{1+e^{-E_{1}^{o}\left(R_s\right)\left(\log_{10} \rho_{b}-E_{2}^{o}\left(R_s\right)\right)}}$ are introduced. Then, we fix $\mathbf{w}_0$ and $\mathbf{w}_c$, and Problem (P4.2) is reformulated as
\begin{align}
	\left(\text{P4.3}\right)\mathop{\text{min}}\limits_{R_c,\nu,\hat{\rho}_b} ~~~~&~10^{\hat{D_{o}^{s}}\left(R_s\right)+e^{\nu}}\label{34}\\
	\text{s.t.} 
	~~~~~~&~U\left(R_c\right)\leq \hat{\rho}_b\label{35}\\
	~~~~&~-e^{\nu-\beta_1}-e^{\nu}+\hat{D}_o^{c}\left(R_s\right)\leq 0\label{36},\\
	~~~~&~(20)\label{37},
\end{align}
where $U\left(R_c\right)=\text{log}_{10}\left(\frac{Q\left(\mathbf{ln}_2{\sqrt{L}\left(C\left(\gamma\right)-R_c\right)}\right)}{R_c L}\right)$,  (\ref{35}) and (\ref{36}) are added non-convex constrains caused by introducing auxiliary variables $\hat{\rho}_b$ and $e^{\nu}$ and $\beta_1=E_1^{o}\left(R_s\right)\left(\hat{\rho}_b-E_2^{o}\left(R_s\right)\right)$. Moreover, (\ref{35}) is derived by approximating $\sqrt{1-\frac{1}{\left(1+\gamma\right)^2}}$ as $1$\cite{yuan2025adaptivesourcechannelcodingmultiuser}. It is evident that (22) monotonically increases with respect to $\hat{\rho}_b$ and $e^{\nu}$. 
Consequently, all constraints in (23)-(25) must hold with equality for the optimal solution to Problem (P4.3); otherwise, (22) can be further reduced by decreasing $\hat{\rho}_b$ and $e^{\nu}$, which confirms the equivalence of the transformation from Problem (P4.3) to Problem (P4.2). To address  non-convex constraints (\ref{35})-(\ref{36}),  the SCA method is employed, which constructs a convex upper bound for the left-hand side (LHS) expressions of the non-convex constraints. Then Problem (P4.4) is rewritten as
\begin{align}
	\left(\text{P4.4}\right)\mathop{\text{min}}\limits_{R_c,\nu,\hat{\rho}_b}~~~~&~10^{\hat{D_{o}^{s}}\left(R_s\right)+e^{\nu}}\label{42}\\
	\text{s.t.} 
	~~~~~~&~U_1\left(R_c^{\left(i\right)},R_c\right)- \hat{\rho}_b\leq 0\label{43},\\
	~~~~&~U_{2}\left(\nu,\nu^{\left(i\right)},\hat{\rho}_b,\hat{\rho}_b^{\left(i\right)}\right)\leq 0\label{44},\\
	~~~~&~(20)\label{45},
\end{align} 
where
\begin{equation}
\begin{aligned}
		U_1\left(R_c^{\left(i\right)},R_c\right)&=\text{log}_{10}Q\left(\hat{	\psi}\right)\\
		&-\text{log}_{10}\left(R_cL\right)-\hat{a}\left(\hat{	\psi}\right)\left(\psi-\hat{	\psi}\right)\text{log}_{10}^{e}\label{40},
	\end{aligned}
\end{equation}
with  $\psi=\sqrt{L}\left(C\left(\frac{\left|\mathbf{h}_c^{H}\mathbf{w}_c^{\left(i\right)}\right|^2}{\left|\mathbf{h}_c^{H}\mathbf{w}_0^{\left(i\right)}\right|^2+\sigma_{n}^2}\right)-R_c\right)\ln 2$ and $\hat{\psi}=\sqrt{L}\left(C\left(\frac{\left|\mathbf{h}_c^{H}\mathbf{w}_c^{\left(i\right)}\right|^2}{\left|\mathbf{h}_c^{H}\mathbf{w}_0^{\left(i\right)}\right|^2+\sigma_{n}^2}\right)-R_c^{\left(i\right)}\right)\ln2$, respectively. Moreover, the specific expression of $U_{2}\left(\nu,\nu^{\left(i\right)},\hat{\rho}_b,\hat{\rho}_b^{\left(i\right)}\right)$ is expanded at (31), with $\mathbf{w}_c^{\left(i\right)}$, $\mathbf{w}_0^{\left(i\right)}$ $R_c^{\left(i\right)},\nu^{\left(i\right)}$ and $\hat{\rho}_{b}^{\left(i\right)}$ denoting the feasible solution obtained from the $i$-th iteration. Then, Problem (P4.4) can be solved by CVX.
\begin{figure*}[b]
	\hrulefill
	\vspace{0in}
	\setcounter{TempEqCnt}{\value{equation}}
	\begin{equation}
		\begin{aligned}
			U_{2}\left(\nu,\nu^{\left(i\right)},\hat{\rho}_b,\hat{\rho}_b^{\left(i\right)}\right)=&-e^{\nu^{\left(i\right)}-E_1^{o}\left(R_s\right)\left(\hat{\rho}_b^{\left(i\right)}-E_2^{o}\left(R_s\right)\right)}\left(\nu-\nu^{\left(i\right)}-E_1^{o}\left(R_s\right)\left(\hat{\rho}_b-\hat{\rho}_b^{\left(i\right)}+1\right)\right)
			-e^{\nu^{\left(i\right)}}\left(\nu-\nu^{\left(i\right)}+1\right)+\hat{D_{o}^{s}}\left(R_s\right)\label{41},
		\end{aligned}
	\end{equation}
	\begin{equation}
		\begin{aligned}
			G_j(\mathbf{w}_c, \mathbf{w}_0) &= 2 \left\| (\mathbf{w}_c^{(i)})^H (\zeta_j) \mathbf{w}_c \right\|^2
			+ 2 \left\| (\mathbf{w}_0^{(i)})^H (\zeta_j) \mathbf{w}_0 \right\|^2 \\
			&\quad + 2\mathcal{R} \left\{ \left( \mathbf{a}_{1,j}^{(i)} - (\mathbf{w}_0^{(i)})^H \mathbf{\Lambda}_j \right) \mathbf{w}_0 \right\}
			+ 2\mathcal{R} \left\{ \left( \mathbf{a}_{2,j}^{(i)} - (\mathbf{w}_c^{(i)})^H \mathbf{\Lambda}_j \right) \mathbf{w}_c \right\},
		\end{aligned}
		\tag{48}
	\end{equation}
	\begin{equation}
		\begin{aligned}
			\mathbf{\Lambda}_j&=\left(\Gamma_{j}\right)^{*}\boldsymbol{\zeta}_j+\Gamma_{j}\left(\boldsymbol{\zeta}_j\right)^{H}, h^{(i)} = (\mathbf{w}_c^{(i)})^H \zeta_j,g^{(i)} = (\mathbf{w}_0^{(i)})^H \zeta_j \mathbf{w}_0^{(i)} \mathbf{w}_c^{(i)},\\
			\mathbf{a}_{1,j}^{\left(i\right)}&=\left(\left(h^{\left(i\right)}\right)^{*}\left(\boldsymbol{\zeta}_j\right)\mathbf{w}_0^{\left(i\right)}+h^{\left(i\right)}\left(\boldsymbol{\zeta}_j\right)^H\mathbf{w}_0^{\left(i\right)}\right)^H,
			\mathbf{a}_{2,j}^{(i)}= \left( (g^{(i)})^* (\zeta_j) \mathbf{w}_c^{(i)} + g^{(i)} (\zeta_j)^H \mathbf{w}_c^{(i)} \right)^H. \\
		\end{aligned}
		\tag{48}
	\end{equation}
\end{figure*}
\subsubsection{Beamforming Optimization}
When the optimal $R_c^{\star}$ is obtained from Problem~(P4.4), (\ref{42}) increases monotonically with $\rho_b$ as shown in~(\ref{13}). Since the BER $\rho_{b}$ decreases monotonically with $\gamma$ according to the first‑order derivative of~(\ref{7}), Problem~(P4.4) is equivalent to maximizing  SINR~$\gamma$, i.e.,
\begin{align}
	\left(\text{P4.5}\right)\mathop{\text{max}}\limits_{\mathbf{w}_0, \mathbf{w}_c} ~~~~&~\frac{\left|\mathbf{h}_c^{H}\mathbf{w}_c\right|^2}{\left|\mathbf{h}_c^{H}\mathbf{w}_0\right|^2+\sigma_{n}^2}\label{50}\\ 
	\text{s.t.} ~~~~~&~(15),(18)\label{51}.
\end{align}	
As  (32) is also non-convex, we leverage  FP method to convert it into a convex version, i.e., 
\begin{align}
 2\mathcal{R}\left(y\mathbf{h}_c^{H}\mathbf{w}_c\right)-\left|y\right|^2\left(\left|\mathbf{h}_c^{H}\mathbf{w}_0\right|^2+\sigma_{n}^2\right)\label{52}
\end{align}
where $ y = \mathbf{w}_c^H \mathbf{h}_c / (|\mathbf{h}_c^H \mathbf{w}_0|^2 + \sigma_n^2)$. As the constraint in (15) is still non-convex, we utilize the positive semi-definite property   of the FIM matrix $\mathbf{J}_{D}^{\mathbf{c}_1}-\mathbf{J}_{D}^{\mathbf{c}_2}\left(\mathbf{J}_{D}^{c_3}+\mathbf{J}_{B}^{c_3}\right)\left(\mathbf{J}_{D}^{\mathbf{c}_2}\right)^T$ and the property that $\text{Tr}\left(\mathbf{U}^{-1}\right)$ is a decreasing function on the positive semi-definite matrix space with a given matrix $\mathbf{U}$ to transform  the non-convex constraint in (\ref{27}) into the following form by introducing an auxiliary matrix $\mathbf{\Omega}\in\mathbb{C}^{2\times2}$, i.e., 
\begin{equation}\label{54}
	\text{Tr}\left(\mathbf{\Omega}^{-1}\right)\leq\Pi, \mathbf{\Omega}\succeq\mathbf{0},
\end{equation}
\begin{equation}\label{55}
\mathbf{J}\left(\boldsymbol{\mu}\right) \succeq \mathbf{\Omega},
\end{equation}  
By utilizing Schur complement, (\ref{55}) is further rewritten as
\begin{equation}\label{56}
	\begin{bmatrix}
		\mathbf{J}_D^{\mathbf{c}_1}-\mathbf{\Omega}&\mathbf{J}_D^{\mathbf{c}_2} \\
		\left(\mathbf{J}_D^{\mathbf{c}_2}\right)^T&\mathbf{J}_D^{c_3}+\mathbf{J}_B^{c_3}
	\end{bmatrix}\succeq \mathbf{0}.
\end{equation}
It is obvious that constraint (\ref{54}) is convex, while  (\ref{56}) is non-convex. Hence, the auxiliary variable vector $\mathbf{\Gamma}=\left[\Gamma_1,\Gamma_2,\Gamma_3,\Gamma_4,\Gamma_5,\Gamma_6\right]^{T}\in\mathbb{C}^{6\times1}$ is introduced to extract $\mathbf{w}_0$ and $\mathbf{w}_c$ from (\ref{56}).  Problem (P4.5) is reformulated as
\begin{align}
	\left(\text{P4.6}\right)\mathop{\text{max}}\limits_{y,\boldsymbol{\Omega}, \mathbf{w}_0, \mathbf{w}_c,\mathbf{\Gamma}} ~&~ (\ref{52})\label{57}\\
	\text{s.t.} ~~~~~&\begin{bmatrix}
		\frac{2}{\sigma_{s}^{2}}\mathcal{R}\left(\mathbf{a}_1\right)-\mathbf{\Omega}&\frac{2}{\sigma_{s}^{2}}\mathcal{R}\left(\mathbf{a}_2\right)\\
		\frac{2}{\sigma_{s}^{2}}\mathcal{R}\left(\mathbf{a}_2^{T}\right)&\frac{2}{\sigma_{s}^{2}}\mathcal{R}\left(\mathbf{a}_3\right)+\frac{1}{\sigma_{\triangle}^2}
	\end{bmatrix}\succeq\mathbf{0}\label{58},\\
	~~~~&~\Gamma_{j}=f_j\left(\mathbf{w}_c, \mathbf{w}_0\right),\label{59}\\
	~~~~&~(18), (35)\label{60},
\end{align}
where $f_j\left(\mathbf{w}_c, \mathbf{w}_0\right)=\text{Tr}\left(\boldsymbol{\zeta}_{j}\boldsymbol{\Sigma}\right)$, $\mathbf{a}_1=\begin{bmatrix}
	\Gamma_1&\Gamma_2\\
	\Gamma_2&\Gamma_3
\end{bmatrix}$, $\mathbf{a}_2=\begin{bmatrix}
	\Gamma_4\\
	\Gamma_5
\end{bmatrix}$, and $\mathbf{a}_3=\Gamma_6$, $\forall j\in\boldsymbol{\mathcal{J}}=\left\{1,2,...,6\right\}$. Let $\mathbf{S}\left(y, \mathbf{w}_0, \mathbf{w}_c\right)=2\mathcal{R}\left(y\mathbf{h}_c^{H}\mathbf{w}_c\right)-\left|y\right|^2\left(\left|\mathbf{h}_c^{H}\mathbf{w}_0\right|^2+\sigma_{n}^2\right)$.
Then, adding non-convex constraint (\ref{59}) as a penalty term to the objective function (\ref{57}), then Problem (P4.6) is reformulated as
\begin{align}
	\left(\text{P4.7}\right)\mathop{\text{min}}\limits_{y,\boldsymbol{\Omega}, \mathbf{w}_c,\mathbf{w}_0, \mathbf{\Gamma}} ~~~~&~-\mathbf{S}\left(y, \mathbf{w}_0, \mathbf{w}_c\right)+\frac{1}{2\kappa}F\left(\mathbf{w}_c, \mathbf{w}_0, \mathbf{\Gamma}\right)\label{61}\\
	\text{s.t.} 
	~~~~~~~~~&~(18),(35),(39)\label{62},
\end{align}  
where $\kappa\textgreater0$ denotes the penalty parameter and $F\left(\mathbf{w}_c,\mathbf{w}_0, \boldsymbol{\Gamma}\right)=\sum_{j=1}^{6}\left|f_j\left(\mathbf{w}_c,\mathbf{w}_0\right)-\Gamma_j\right|^2$. To solve Problem (P4.7), an AO method is utilized. Specifically, we first fix the beamforming vectors $\mathbf{w}_c$ and $\mathbf{w}_0$, the sub-problem for acquiring $\boldsymbol{\Omega}$ and $\boldsymbol{\Gamma}$ is written as
\begin{align}
	\left(\text{P4.8}\right)\mathop{\text{min}}\limits_{\boldsymbol{\Omega},  \mathbf{\Gamma}}~~~~&~\frac{1}{2\kappa}F\left(\mathbf{w}_c,\mathbf{w}_0,\mathbf{\Gamma}\right)\label{63},\\
	\text{s.t.} 
	~~~~&~(35),(39)\label{64},
\end{align} 
which is a semi-definite programming problem (SDP) and can be solved by CVX. Similarly, by adopting the SCA method, the sub-problem for $\mathbf{w}_0$ and $\mathbf{w}_c$ is reconstructed as
\begin{equation}\label{68}
	\begin{aligned}
		\left(\text{P4.9}\right)\mathop{\text{min}}\limits_{y,\mathbf{w}_0,\mathbf{w}_c  }~~~~&~F_1\left(y, \mathbf{w}_c,\mathbf{w}_0, \boldsymbol{\Gamma}\right),\\
		\text{s.t.} 
		~~~~~~&~(18),
	\end{aligned}	
\end{equation}
where  
$\kappa=\iota\cdot\kappa^{i}$ and $F_1\left(\mathbf{w}_c,\mathbf{w}_0, \boldsymbol{\Gamma}\right)=-\mathbf{S}\left(y, \mathbf{w}_0, \mathbf{w}_c\right)+\sum_{j=1}^{6}\frac{1}{2\kappa}G_j\left(\mathbf{w}_c, \mathbf{w}_0\right)$, with $G_j\left(\mathbf{w}_c, \mathbf{w}_0\right)$ given in (48)-(49), and $\iota$ denoting the step size to control the decreasing speed of $\kappa$.
Obviously, Problem (4.9) is a convex problem, which can be solved by  CVX. 
\section{Simulation Results}
We consider the ISSC system with one TR equipped with $N_T=4$ antennas, one target, one single-antenna CRE and one SRE with $N_R=4$ antennas. The positions of the BS, target, SRE and CRE are set as $\mathbf{p}_b=\left[0,0\right], \mathbf{p}_o=\left[100,50\right], \mathbf{p}_r=\left[0,50\right], \mathbf{p}_s=\left[20,30\right]$, respectively.   
Other parameters are set as $\epsilon =3$, $\sigma_n^2=\sigma_s^2=-50 \text {dBm}$, $\sigma_\Delta=100$ns, $\beta_s=0.6, L=256, B=100$MHz, $ N=1024$. The comparison benchmark is selected as DJSCC–water‑filling (WF)–zero‑forcing (ZF), where DJSCC is employed to extract feature information, and the WF algorithm is applied to allocate power for ZF beamforming. Besides, multi-Scale structural similarity (MS-SSIM) is selected as the performance metric of SemCom  \cite{yuan2025adaptivesourcechannelcodingmultiuser}.
\vspace{2mm} 
\noindent
\begin{minipage}[t]{0.48\columnwidth}
	\centering
	\includegraphics[width=\linewidth]{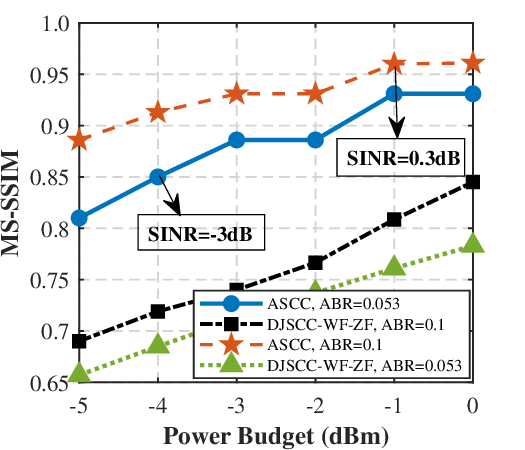}
	\captionof{figure}{MS-SSIM comparison under different schemes}
	\label{fig:psnr}
\end{minipage}
\hfill
\begin{minipage}[t]{0.48\columnwidth}
	\centering
	\includegraphics[width=\linewidth]{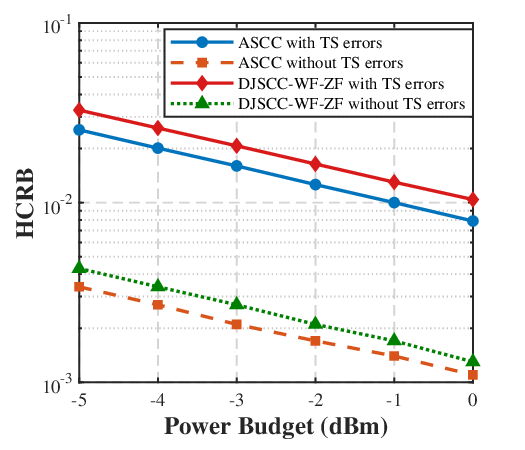}
	\captionof{figure}{HCRB comparison under different schemes}
	\label{fig:crb}
\end{minipage}

As shown in Fig.~\ref{fig:psnr}, the MS-SSIM of the proposed ASCC scheme consistently surpasses the DJSCC-WF-ZF benchmark across various $P_T$ and ABR settings, benefiting from its adaptive model-selection mechanism that selects the optimal encoder–decoder pair from $G$ pre-trained models based on channel conditions.  
Even with low ABR, the scheme maintains favorable image reconstruction performance at $\mathrm{SINR} = -3\,\mathrm{dB}$, demonstrating the advantage of semantic communications in low-SNR scenarios.  
The performance curves exhibit plateaus at certain power points because the selected model has nearly reached its limit at the preceding level, and a $1\,\mathrm{dBm}$ increase in power is insufficient to raise the SINR beyond the threshold for switching to the next model with higher $R_s$.

As shown in Fig.~\ref{fig:crb}, the HCRB of the proposed ASCC scheme outperforms the DJSCC-WF‑ZF benchmark  across different $P_T$. This improvement can be attributed to the ability of our scheme to maintain reasonable semantic  performance under the low SINR region, enabling more power for sensing. These results validate the effectiveness of ISSC system. Furthermore, the notable degradation caused by TS errors underscores the need to account for them in system design.

 %However, the $K$-means method can get a larger communication rate threshold, which is due to the fact that more power is allocated for communications and less for sensing in $K$-means based user selection schemes to achieve a higher CRB.
 
% 
% the communication channel conditions between the BS and corresponding ISAC receivers are better than that of the proposed minimax method. 
   ha

\section{Conclusion}
This paper proposed an ASCC scheme for ISSC system under imperfect TS. The E2E semantic distortion was  derived in closed form via the data regression method, and positioning performance with TS errors was evaluated by the HCRB. An E2E distortion minimization problem was then formulated by jointly designing the coding rate and  beamforming. An  AO algorithm composed of SCA and FP was developed to solve this problem, and simulations validated our analysis.

%
%that the proposed MATSPG algorithm has significant advantages in the problems with large discrete action spaces compared with the traditional DRL algorithms.	

% 
\def\baselinestretch{0.7}
\bibliography{reference2}
\end{document}